# HEAT TRANSPORT THROUGH PLASMONIC INTERACTIONS IN CLOSELY SPACED METALLIC NANOPARTICLES CHAINS

Philippe Ben-Abdallah[1*], Karl Joulain[2], Jérémie Drevillon[1] and Clément Le Goff[1]

[1]Laboratoire de Thermocinétique, CNRS UMR 6607, Ecole Polytechnique de l'Université de Nantes, 44 306 Nantes cedex 03, France.

[2]Laboratoire d'Études Thermiques (LET)-ENSMA, 1 Avenue Clément Ader, BP 40109, 86961 Futuroscope Chasseneuil Cedex, France.

* pba@univ-nantes.fr

**Abstract**

We report a numerical investigation on the heat transfer through one dimensional arrays of metallic nanoparticles closely spaced in a host material. Our simulations show that the multipolar interactions play a crucial role in the heat transport via collective plasmons. Calculations of the plasmonic thermal conductance and of the thermal conductivity in ballistic and diffusive regime, respectively have been carried out. (a) Using the Landauer-Buttiker formalism we have found that, when the host material dielectric constant takes positive values, the multipolar interactions drastically enhance by several order of magnitude the ballistic thermal conductance of collective plasmons compared with that of a classical dipolar chain. On the contrary, when the host material dielectric constant takes negative values, we have demonstrated the existence of non-ballistic multipolar modes which annihilate the heat transfer through the chains. (b) Using the kinetic theory we have also examined the thermal behavior of chains in the diffusion approximation. We have shown that the plasmonic thermal conductivity of metallic nanoparticle chains can reach 1% of the bulk metal thermal conductivity . This result could explain the anomalously high thermal conductivity observed in many colloïdal suspensions, the so called nanofluids.

## I. INTRODUCTION

When the separation distance $d$ between two polarized or charged nano-objects decreases, the magnitude of their electrostatic interactions [1] increases as well as the heat and momentum they exchanges. When $d$ is much larger than the nano-objects characteristic size $D$, the energy they exchange is mainly due to their dipolar interactions [2]. On the contrary, at smallest separation distances, the multipolar fields become dominant compare to the dipolar ones and the energy of multipolar interactions increase with the decreasing distance $d$ as $O(d^{-(2l+1)})$, $l$ denoting the multipole order (for dipole interactions $l=1$). These interactions lead to a very strong enhancement of exchanges close to the contact [3-5].

In closely spaced nano-objects lattices (i.e. $d<2D$), the physics of heat and momentum transfer is more complex and still remains to be explained. Indeed, in such structures, the multipolar moments supported by each object are able to interact with their counterparts giving rise to new channels for heat and momentum transport through the so called "collective plasmon modes" (CPM). In a recent work [6], we have studied the heat transport through metallic nanoparticle chains by considering only the CPM due to dipolar interactions. Although we have shown that the thermal conductance of such chains was greater by at least six orders of magnitude than the far-field thermal conductance ($d>>D$) between two particles, we have seen than this conductance was far below the conductance of bulk metal.

In the present work, we show that the physics radically changes when the separation distance between two neighboring particles is small compared to their characteristic size ($d<<2D$). We will see that, under these conditions, the multipolar interactions play a major role in the heat transport and can give rise to new optical and thermal behaviors. The thermal behavior of chains associated to CPM will be studied using both the Landauer-Buttiker theory [7] for the ballistic regime and using the kinetic theory [8] for diffusive regime, respectively.

## II. DISPERSION RELATION OF COLLECTIVE PLASMONS MODES

Let us consider a one-dimensional straight chain of polarizable metallic spherical particles dispersed in a homogeneous isotropic dielectric host material. Thermal fluctuations in this composite medium lead to a displacement of electric charges in each nanoparticle and consequently give rise to local multipolar moment. An electric field applied on the charge distribution of the $n$ th particle induces on it a multipolar moment of order $(l,m)$ [9]

$$q_{lmn} = -\frac{2l+1}{4\pi}\alpha_{lmn}V_{lm}(n), \qquad (1)$$

where $V_{lm}(n)$ is the coefficient of order $(l,m)$ of the expansion of local electric potential while $\alpha_{lmn}$ is the polarizability of the (spherical) particle which is defined by [10-11]

$$\alpha_{lmn} = \frac{l(\varepsilon_p - \varepsilon_m)}{l(\varepsilon_p + \varepsilon_m)+1} a^{2l+1}, \qquad (2)$$

$\varepsilon_p$ being the dielectric permittivity of particle, $\varepsilon_m$ that of host medium and $a$ the radius of nanoparticles (Fig.1). Then, the electric potential at the position of the $n$ th particle takes the present form

$$V_{lm}(n) = V_{lm}^{ext}(n) + \sum_{n' \neq n} V_{lm}^{n'}(n), \qquad (3)$$

where $V_{lm}^{n'}(n)$ is the electric potential at position of the $n$ th particle from the $n'$ th particle. Expressing this expression in terms of multipolar moments we obtain

$$q_{lmn} = -\frac{2l+1}{4\pi}\alpha_{lmn}[V_{lm}^{ext} + \sum_{\substack{l'm' \\ n \neq n'}} (-1)^{l'} A_{lmn}^{l'm'n'} q_{l'm'n'}], \qquad (4)$$

where $A_{lmn}^{l'm'n'}$ is the coupling strength between the multipole moment of order $(l,m)$ at the position of the $n$ th particle and the multipole moment of order $(l',m')$ at the $n'$ th particle. According to the theory of spectral representation introduced by [12] this coefficient is purely geometric and writes, for one-dimensional periodic lattices

$$A_{lmn}^{l'm'n'} = \begin{cases} 0 & \text{if } n = n' \\ (-1)^m 4\pi \dfrac{(l+l')!}{[(2l+1)(2l'+1)]^{1/2}} \dfrac{1}{[(l+m)!(l'+m)!(l-m)!(l'-m)!]^{1/2}} \\ \quad \times (\dfrac{n'-n}{|n'-n|})^{l+l'} \dfrac{1}{d^{l+l'+1}|n'-n|^{l+l'+1}} \delta_m^{m'} & \text{if } n \neq n' \end{cases} \qquad (5)$$

Notice that when the size $L$ of the chain is smaller than the characteristic wavelength $\lambda = 2\pi c / \alpha \omega_p$ (on Figs. 2 and 4 we verify that $\alpha < 1$) of electromagnetic waves which propagate along it, we can apply the electrostatic approximation to evaluate heat exchanges due to electromagnetic transport through the medium. This condition will be always fulfilled in the simulations carried out in the present work.

Seeking a solution of linear system (4) under a plane wave form

$$q_{nlm} = q_{lm} \exp[j(\omega_{lm} t - knd)], \qquad (6)$$

where $\omega_{lm} = \omega'_{lm} + j\omega''_{lm}$ is a complex pulsation. Using, thanks to the periodicity of problem, the Floquet theorem, we obtain a relation the multipolar moments must satisfy

$$\sum_{l'm'} \left\{ \dfrac{4\pi}{2l'+1} \alpha_{l'm'n}^{-1} \delta_{lm}^{l'm'} + \sum_{n' \neq n} (-1)^{l'} A_{lmn}^{l'm'n'} \exp[j(n-n')kd] \right\} q_{l'm'} = -V_{lm}^{ext} \exp(jnkd), \qquad (7)$$

where $\delta_{lm}^{l'm'}$ is the product of two Kronecker symbols relative to the couples $(l,l')$ and $(m,m')$, respectively. By assuming, that the dielectric constant of nanoparticles is described by the Drude model

$$\varepsilon_p(\omega) = 1 - \dfrac{\omega_p^2}{\omega(\omega - j\gamma)}$$ where $\omega_p$ is the plasma resonance pulsation of metal and $\gamma = \tau^{-1}$ the damping factor (equal to the inverse of the average time between two subsequent electron collisions), the polarizability of order $(l,m)$ of the $n$ th particle writes

$$\alpha_{lmn} = \dfrac{\xi_{lm} t - l}{\xi_{lm} s - l} a^{2l+1}, \qquad (8)$$

where we have set $\xi_{lm} = \omega_{lm}(\omega_{lm} - i\gamma)/\omega_p^2$, $s = l(1+\varepsilon_m)+1$ and $t = l(1-\varepsilon_m)$. Thus, after a straightforward calculation it is easy to show that, the system of equation (7) can be rewritten in the form

$$\sum_{l'm'}([\delta_{lm}^{l'm'} + \frac{1}{4\pi}(2l+1)\frac{t}{s}a^{2l+1}\sum_{n'\neq n}(-1)^{l'} A_{lmn}^{l'm'n'} \exp[j(n-n')kd]]\xi_{l'm'}$$
$$-\frac{l}{s}\delta_{lm}^{l'm'} - \frac{1}{4\pi}\frac{(2l+1)l}{s}a^{2l+1}\sum_{n'\neq n}(-1)^{l'} A_{lmn}^{l'm'n'} \exp[j(n-n')kd])q_{l'm'}$$
$$= -\frac{1}{4\pi}V_{lm}^{ext}\exp(jnkd)\frac{2l+1}{s}[\xi_{lm} t - l]a^{2l+1}$$  (9)

By setting $\overline{q}_{lm} = \dfrac{q_{lm}}{(la^{2l+1})^{1/2}}$ in the previous expression, we obtain the more convenient system

$$\sum_{l'm'}([\frac{s}{2l+1}\delta_{lm}^{l'm'} + \frac{1}{4\pi}\frac{t}{l^{1/2}}l'^{1/2}a^{l+l'+1}\sum_{n'\neq n}(-1)^{l'} A_{lmn}^{l'm'n'} \exp[j(n-n')kd]]\xi_{l'm'}$$
$$-\frac{l}{2l+1}\delta_{lm}^{l'm'} - \frac{1}{4\pi}l'^{1/2}l^{1/2}a^{l+l'+1}\sum_{n'\neq n}(-1)^{l'} A_{lmn}^{l'm'n'} \exp[j(n-n')kd])\overline{q}_{l'm'}$$
$$= -\frac{1}{4\pi}V_{lm}^{ext}\exp(jnkd)[\xi_{lm} t - l]a^{2l+1}$$  (10)

The calculation of eigenvalues $\xi_{lm}$ of this system allows us to determine the dispersion relation of resonant collective modes (eigenmodes) of the chain

$$\omega_{lm}(k) = \frac{1}{2}[\sqrt{4\xi_{lm}(k)\omega_p^2 - \gamma^2} + j\gamma]$$  (11)

According to a well known result of linear algebra, we know that the eigenvalues $\xi_{lm}$ are also solution of the secular equation

$$\det([\frac{s}{2l+1}\overline{\overline{1}} + \frac{t}{l}\overline{\overline{J}}]^{-1}[\frac{l}{2l+1}\overline{\overline{1}} + \overline{\overline{J}}] - \xi_{lm}\overline{\overline{1}}) = 0,$$  (12)

where, the two matrix $\overline{\overline{J}}$ and $\overline{\overline{1}}$ are defined by

$$(\overline{\overline{J}})_{lm}^{l'm'} = \frac{1}{4\pi}l^{1/2}l'^{1/2}a^{l+l'+1}\sum_{n'\neq n}(-1)^{l'} A_{lmn}^{l'm'n'} \exp[j(n-n')kd]$$  (13)

and

$$(\overline{\overline{1}})_{lm}^{l'm'} = \delta_l^{l'}\delta_m^{m'},$$  (14)

$\delta_i^j$ being the usual Kronecker symbol. Here, let us note (see appendix) that the matrix $\bar{\bar{X}} = [\frac{s}{2l+1}\bar{\bar{1}} + \frac{t}{l}\bar{\bar{J}}]^{-1}[\frac{l}{2l+1}\bar{\bar{1}} + \bar{\bar{J}}]$ is a Hermitian matrix so that its eigenvalues $\xi$ are all purely real. Therefore, if the expression under the squareroot symbol in (11) is positive, the collective plasmons are propagative modes with a finite lifetime $\tau = \mathfrak{I}^{-1}(\omega_{lm}) = \gamma^{-1}/2$ (independent on k). When it is neagative, the mode pulsation is imaginary and the corresponding mode does not participate anymore to the ballistic transport through the chain. In the next section we will see that this lead to two radically distinct thermal behaviors.

## II. MULTIPOLAR THERMAL CONDUCTANCE IN BALLISTIC REGIME

### A. Landauer formalism

In order to estimate the multipolar thermal conductance of a chain, we use an approach analogous to that used in the Landauer theory of electronic transport. Two heat reservoirs (Fig.1) which are maintained at neighboring temperatures $T$ and $T + \delta T$ (the temperature difference $\delta T$ being small compared to the mean temperature $\frac{2T + \delta T}{2}$) are linked together by the chain we want to characterize. We assume these reservoirs support surface modes which are able to perfectly couple with the collective plasmon modes of the chain. In this case, the heat flux exchanged between both reservoirs through a chain of length $L$ is given by

$$\varphi^{\pm} = \frac{1}{2\pi} \sum_{lm} \int_0^{\infty} |v_{glm}|(k)\hbar\omega_{lm}(k) f_B[\omega_{lm}(k)] dk, \quad (15)$$

where the signs + and – label the right and left moving flux respectively. In this expression $k$ stand for the wave vector, $\omega_{lm}(k)$ is the pulsation of $m$ th mode and $f_B(\omega) = [\exp(\beta\hbar\omega) - 1]^{-1}$ is the Bose-Einstein distribution function (plasmons are bosons). The thermal conductance of a nanoparticles chain associated to heat transport from plasmons is given by

$$G = \lim_{\delta T \to 0} \frac{\varphi^+(T + \delta T) - \varphi^-(T)}{\delta T}. \quad (16)$$

Finally, performing the Eq. (15) calculation in the first Brillouin zone $[0, \pi/d]$ we obtain the following expression for the conductance

$$G = \frac{\hbar^2}{2\pi k_B T^2} \sum_{lm} \int_0^{\pi/d} \omega_{lm}^2(k) v_{glm}(k) \frac{e^{\beta\hbar\omega_{lm}}}{(e^{\beta\hbar\omega_{lm}} - 1)^2} dk \ . \tag{17}$$

In our study $T = 900K$ and $\omega_p \approx 1.6 \times 10^{16} \, rad.s^{-1}$ (copper), so that the thermal energy $k_B T \approx 1.2 \times 10^{-20} J$ is negligible compare to the waves energy $\hbar\omega_p = 1.6 \times 10^{-18} J$. More generally, a direct inspection of dispersions curves (Figs. 2 and 4) shows that the low frequency modes are not below $0.1\omega_p$ so that for any modes we have $k_B T \ll \hbar\omega_{lm}$. The chain modes at frequencies near $\omega_p$ are not highly populated by thermal excitation but they are the ones of lowest energy. Thus, using the Wien approximation of the Bose-Einstein distribution function, the thermal conductance can be rewritten on the simplified form

$$G = \frac{\hbar^2}{2\pi k_B T^2} \sum_{lm} \int_0^{\pi/d} \omega_{lm}^2(k) v_{glm}(k) \exp[-\beta\hbar\omega_{lm}] dk \ . \tag{17-bis}$$

Since the dispersion curves of low frequency modes are the most distorted curves, the corresponding modes are the quickest in the chain. Then, from the simplified expression of *G*, it clearly appears that these modes (which always are thermally excited) are the main contributors to the heat ballistic transfer. For practical applications it would be interesting to use as nanoparticles materials whose characteristic energy $\hbar\omega_p$ is close to the Wien energy $\hbar\omega_W = 2.821 k_B T$ correspnding to the maximum of the Planck function.

### B. Numerical results

In Fig. 2, we have plotted the dispersion relation of collective plasmons supported by a chain of copper nanoparticles in vacuum when the nanoparticles (10nm radius) are in mechanical contact and separated by a distance $d = 3a$, respectively. The plasma frequency, is given by the usual relation $\omega_p = (\rho_e e^2 / \varepsilon_0 m^*)^{1/2}$ where $\rho_e$, $e$, $\varepsilon_0$ and $m^*$ denote the electronic density, the charge of electrons, the permittivity of vacuum and the effective mass of electrons. The physical properties of copper used to calculate the transport properties were $\rho_e = 8,5 \times 10^{28} m^{-3}$ [13], $\gamma = 1,38 \times 10^{13} s^{-1}$ [14], $m^* = 1,42 \times m_e$ [15] ($m_e$ being the mass of free electrons). In the $G(d)$ curves displayed in Fig. 3, it clearly appears that the higher multipoles ($l>1$) participate to the energy transport only at very short distance. In our calculations, we have used the dispersion relations of periodic (infinite) chains instead of finite chains. This assumption has been verified by Weber and Ford [16] in dipolar chains (in the quasi-static approximation) with a ten of particles. For separation distance $d$ greater than the critical value $\sim 3a$, the branches of quadrupolar modes ($l=2$) and higher modes are flat so that the group velocity of corresponding modes vanishes. For such separation distances, the classical dispersion relation of dipolar chains [17] are recovered. On the contrary, as $d$ is reduced, the branches of multipoles are distorted so that the group velocity of these modes increases and transport heat through the chain. In addition, when the separation distance decreases, the number of contributing multipole increases. This is an argument in favor of an increase of the thermal conductance. However, as shown in the inset of Fig. 3, a saturation mechanism limits the chain thermal conductance to a maximum value ($\approx 10^{-18} W.K^{-1}$) when the multipole of order higher than $l>8$ are incorporated into the calculation. This is simply due to the decrease of the group velocity of higher modes with the mode order and also to the decrease of factor $(\frac{a}{d})^{l+l'+1}$ which appears in the expression of tensor $\overline{\overline{X}}$ with the mode order. This last point tends to vanish the eigenmodes real part so that they do not participate anymore to heat transport.

We have also done calculations of the plasmonic thermal conductance when the nanoparticles are dispersed in a lossless and non dispersing host material. Two distinct configurations have been studied in accordance with the value of the dielectric constant of host material. In the first one $\varepsilon_m = 10$ so that all multipoles are propagative (Fig. 4) and they participate to the ballistic transport of heat. In this case, the thermal behavior of the chain is very similar (Fig.5) to that of a chain in vacuum. Here again, we observe (Fig. 5) both a saturation mechanism near the contact which limits the contribution of multipoles to the modes of lower order ($l<5$) and a monotonic decreasing of the thermal conductance with respect to the separation distance. In the second configuration, the nanoparticles are embedded in a host material with a negative dielectric constant $\varepsilon_m = -2$. In this case, as shown in Fig. 6, the optical behavior of the chain radically changes. Indeed, contrary to the previous configuration, some of multipoles have a purely imaginary pulsation at least over a part of the Brillouin zone. Then, these modes do not participate anymore to the ballistic transport through the chain and they limit the thermal conductance of the chain to very low values. As a direct consequence of the presence of 'non-ballistic' modes, we have found (not plotted) that the thermal conductance drastically falls down to zero in negative dielectric host materials.

The non-ballistic modes also exist in dipolar chains in which they can be described in details (Fig. 6). Indeed, in this latter case the Hamiltonian matrix $\bar{\bar{X}}$ is a $3 \times 3$ diagonal matrix with elements

$$X_{11}(k) = X_{33}(k) = \xi_{1\pm 1} = \frac{1 + 2(\frac{a}{d})^3 \sum_{n>0} \frac{Cos(nkd)}{n^3}}{2 + \varepsilon_m + 2(1 - \varepsilon_m)(\frac{a}{d})^3 \sum_{n>0} \frac{Cos(nkd)}{n^3}} \qquad (18\text{-a})$$

and

$$X_{22}(k) = \xi_{10} = \frac{1 - 4(\frac{a}{d})^3 \sum_{n>0} \frac{Cos(nkd)}{n^3}}{2 + \varepsilon_m - 4(1 - \varepsilon_m)(\frac{a}{d})^3 \sum_{n>0} \frac{Cos(nkd)}{n^3}}, \qquad (18\text{-b})$$

which are relevant to the transversal (T) (two times degenerated) and to the longitudinal (L) modes, respectively. When $4X_{ii}\omega_p^2 - \gamma^2 < 0$ the pulsation of $i$ th mode becomes purely imaginary so that it does not contribute anymore to the heat ballistic transport. As displayed in (Fig.6), this precisely occurs both for the longitudinal and transversal modes near $d/2a = 1$ over different parts of the Brillouin zone. Therefore, when $k$ belongs to the first half part of the Brillouin zone only the two transversal modes participate to the ballistic transport. Conversely, on the second half part of the Brillouin zone the longitudinal mode is the only one which contributes to the heat transport. For large separation distances ($d/2a \approx 1.35$) all modes become again propagative and we recover the classical behavior of dipolar chains (as displayed in Fig. 4 when $l=1$).

## III. MULTIPOLAR THERMAL CONDUCTIVITY IN DIFFUSIVE REGIME

We are now going to examine the second asymptotic regime of the chain that is its Fourier regime. In this region the heat carriers undergo a lot of collision events and the heat flow can be considered in the diffusion approximation. In order to calculate the thermal conductance in this regime we proceed as follow. First, we calculate the heat flux at a given position $z$ when the chain is submitted to a temperature gradient. During a time interval $dt$, the heat flux is the ratio of the energy balance due to the heat carriers coming from the positions lower than and larger than $z$ over $dt$. On average, the carriers coming from the positions lower than than $z$ had their last collision at the position $z - \ell_{lm}$, $\ell = v_{glm} \tau_{lm}$ being the mean free path of carriers. In the same way, the carriers coming from the positions larger than $z$ had their last collision event at the position $z + \ell_{lm}$. Therefore, the heat flux across the chain sectional surface $S = \pi a^2$ is

$$j_{lm} = [n(z - \ell_{lm})E(z - \ell_{lm}) - n(z + \ell_{lm})E(z + \ell_{lm})]\frac{v_{glm}}{S} = -2\ell_{lm}\frac{v_{glm}}{S}\frac{\partial nE}{\partial z}, \tag{19}$$

$n(z)$ being the carrier density per unit length and $E(z)$ the energy of carriers at position $z$. As the one dimension density of states is $\rho(k)dk = \frac{Ldk}{\pi}$, the product $nE$ writes

$$nE = \frac{\rho(k)dk}{L}\hbar\omega_{lm}(k)f_B[\omega_{lm}(k),T]. \tag{20}$$

In this expression only the Boltzmann function depends on $z$ through the temperature field $T$. Then the heat flux per mode at the wavenumber $k$ writes

$$j_{lm}dk = -2\frac{v_{glm}^2(k)}{S}\tau_{lm}(k)k_B(\frac{\hbar\omega_{lm}(k)}{k_BT})^2 \exp(\frac{\hbar\omega_{lm}}{k_BT})f_B^2[\omega_{lm}(k),T]\frac{dk}{\pi}\frac{\partial T}{\partial z}. \tag{21}$$

We recognize here the expression of the Fourier law at a given wavenumber. After integration over the wavenumbers and after summation over all the branches of spectrum, the thermal conductivity in diffusive regime reads

$$\kappa = \frac{2}{\pi S}\frac{\hbar^2}{k_BT^2}\sum_{lm}\int_0^{\pi/d} v_{glm}^2(k)\tau_{lm}(k)\omega_{lm}^2(k)\frac{e^{\beta\hbar\omega_{lm}}}{(e^{\beta\hbar\omega_{lm}}-1)^2}dk. \tag{22}$$

Let us note that, in this regime, it is more pertinent to consider the thermal conductivity rather than the thermal conductance since the latter depends on the chain length $L$ ($G \infty L^{-1}$) and consequently is not an intrinsic property of medium. Fig. 7 and Fig. 8-b shows the diffusive thermal conductivity of nanoparticle chains versus the separation distance between nanoparticles and versus the dielectric constant of host material for a copper chain constituted of one thousand nanoparticles. It can be seen in Figs. 7 and 8 that $\kappa(d)$ and $\kappa(\varepsilon_m)$ are monotonic curves, the highest values of $\kappa$ being obtained in contact ($d=2a$) and for the highest values of the dielectric constant of host material. A direct inspection of expression (22) seems to show that $\kappa$ increases as $a^{-2}$. However, in Fig. 8-c we see that, contrary to this trivial reasoning which would lead to a plasmonic thermal conductivity for a chain made with copper nanoparticles 1nm diameter of about 40 W.m$^{-1}$.K$^{-1}$, reach in fact a maximum value of 1.6 W.m$^{-1}$.K$^{-1}$ that is less than 0.5% of the bulk thermal conductivity of copper ($\kappa_{Bulk} = 3.72 \times 10^2 W.m^{-1}.K^{-1}$ at 900 K). This result is simply due to the strong dependence of the dispersion relation of collective plasmons which tends to decrease the group velocity of modes as the size of particles decreases.

## IV. CONCLUSIONS

We have studied the heat transport through collective plasmons modes in closely separated nanoparticles chains both in ballistic and Fourier regime. For both regimes, we have demonstrated that there is a strong deviation between the thermal behavior of multipolar chains and that of dipolar chains. In ballistic regime two distinct behaviors have been highlighted. First, when the nanoparticles are dispersed in a dielectric host which dielectric constant is positive, the multipole interactions enhance, by several order of magnitude, the thermal conductance of the chain near the contact. On the contrary, in a dielectric matrix which dielectric constant is negative, some of the modes become non-propagative and do not participate anymore to the heat exchanges so that the thermal conductance in this case becomes extremely low.

In diffusive regime, we have shown that the plasmonic thermal conductivity of long chains of nanosized metallic particles in contact can reach 1% of the thermal conductivity of bulk metal. In interconnected three dimensional lattices of metallic particles such as clusters one can expect that the multiplicity of coupling channels between the plasmons enhances more again the plasmonic thermal conductivity.

**Appendix A:**

Here we prove that the tensor J defined in (12) can be recast into the form

$$J_{lm}^{l'm'} = \begin{cases} 2\sum_{n>0} \frac{Cos(nkd)}{n^{l+l'+1}} U_{ll'm} \delta_{m}^{m'} , & l+l' \text{ even} \\ 2j\sum_{n>0} \frac{Sin(nkd)}{n^{l+l'+1}} U_{ll'm} \delta_{m}^{m'} , & l+l' \text{ odd} \end{cases} \quad (A-1)$$

with

$$U_{ll'm} = (-1)^{m+l} (\frac{a}{d})^{l+l'+1} [\frac{ll'}{(2l+1)(2l'+1)}]^{1/2} \times \frac{(l+l')!}{[(l+m)!(l'+m)!(l-m)!(l'-m)!]^{1/2}} \quad (A-2)$$

and this tensor has purely real eigenvalue.

To prove this assertion let us start by setting the matrix

$$J_{lm}^{l'm'} = \sum_{n' \neq n} A_{nlm,n'l'm'} \exp[j(n-n')kd]. \quad (A-3)$$

This matrix can equivalently be rewritten under the form

$$J_{lm}^{l'm'} = \sum_{n>0} A_{0lm,-nl'm'} \exp[jnkd] + \sum_{n>0} A_{0lm,nl'm'} \exp[-jnkd]. \quad (A-4)$$

Since

$$A_{0lm,-nl'm'} = (-1)^{l+l'} A_{0lm,nl'm'} \quad (A-5)$$

we have also

$$J_{lm}^{l'm'} = \sum_{n>0} A_{0lm,nl'm'} [\exp(jnkd) + (-1)^{l+l'} \exp(-jnkd)] \quad (A-6)$$

Then, by introducing the trigonometric functions we see that

$$J_{lm}^{l'm'} = \begin{cases} 2\sum_{n>0} A_{0lm,nl'm'} Cos(nkd) , & l+l' \text{ even} \\ 2j\sum_{n>0} A_{0lm,nl'm'} Sin(nkd) , & l+l' \text{ odd} \end{cases}. \quad (A-7)$$

But according to relation (5)

$$A_{0lm,nl'm'} = (-1)^{m+l+l'} 4\pi (\frac{1}{nd})^{l+l'+1} \\ \times \frac{1}{[(2l+1)(2l'+1)]^{1/2}} \times \frac{(l+l')!}{[(l+m)!(l'+m)!(l-m)!(l'-m)!]^{1/2}} \times \delta_{m}^{m'} . \quad (A-8)$$

Thus,

$$J_{lm}^{l'm'} = \begin{cases} 2\sum_{n>0} \frac{Cos(nkd)}{n^{l+l'+1}} U_{ll'm} \delta_m^{m'}, & l+l' \text{ even} \\ 2j\sum_{n>0} \frac{Cos(nkd)}{n^{l+l'+1}} U_{ll'm} \delta_m^{m'}, & l+l' \text{ odd} \end{cases}, \qquad (A-9)$$

which prove our assertion. Moreover since

$$U_{ll'm} = \begin{cases} U_{l'lm}, & \text{if } l+l' \text{ even} \\ -U_{l'lm}, & \text{if } l+l' \text{ odd} \end{cases} \qquad (A-10)$$

the complex conjugate $\bar{J}$ of tensor $J$ reads

$$\bar{J}_{lm}^{l'm'} = \begin{cases} J_{l'm'}^{lm} & \text{when } l+l' \text{ even} \\ -2j\sum_{n>0} \frac{Sin(nkd)}{n^{l+l'+1}} U_{ll'm} \delta_m^{m'} = 2j\sum_{n>0} \frac{Sin(nkd)}{n^{l+l'+1}} U_{l'lm} \delta_m^{m'} = J_{l'm'}^{lm} & \text{when } l+l' \text{ odd} \end{cases} \qquad (A-11)$$

so that J is an Hermitian tensor and this result achieved the demonstration of the second part of our assertion. Note that it is straightforward to see that we have also as immediate consequence

$$\bar{X}_{lm}^{l'm'} = X_{l'm'}^{lm}. \qquad (A-12)$$

**Figure captions**

Fig. 1 : Regularly spaced nanoparticles chain connected to two reservoirs kept at ambient temperatures. The separation distance $d$ between particles is smaller than the wavelength of plasmons modes supported by the chain.

Fig. 2 : Dispersion relation of collective plasmons modes along a chain of 10 nm radius copper particles dispersed in vacuum in contact (black curves) and spaced by 30 nm (blue curves). The curves are computed by incorporating : the dipolar moment ($l$=1), the quadrupolar moments ($l$=2), all the multipolar moments up to  $l$=5 and $l$=8.

Fig. 3 :  Plasmonic thermal conductance at 900 K of a linear chains of 10 copper particles 10 nm radius in vacuum versus the separation distance. The curves are computed by including : the dipolar moments ($l$=1), all the multipoles up to $l$=3,  $l$=5 and $l$=8. The inset is a zoom on the near contact region.

Fig. 4 : Dispersion relations of collective plasmons modes along a chain of 10 nm radius copper particles in mechanical contact and dispersed in a host dielectric ($\varepsilon_m = 10$) by incorporating : the dipolar moments ($l=1$), all the multipoles up to $l=3$ and up to $l=5$.

Fig. 5 : Plasmonic thermal conductance at 900 K of a linear chains of 10 copper particles 10 nm radius (ballistic regime) in vacuum versus the separation distance. The curves are computed including : the dipolar moments ($l=1$), all the multipoles up to $l=3$ and $l=5$.

Fig. 6 : Dispersion relations of collective plasmons modes along a chain of 10 nm radius copper particles dispersed in a negative dielectric ($\varepsilon_m = -2$) by incorporating : the dipolar moments ($l=1$), all the multipoles up to $l=3$ and up to $l=5$.

Fig. 7 : Plasmonic thermal conductivity at 900 K of a linear chains of 1000 copper particles 10 nm radius (diffusive regime) dispersed in a host dielectric ($\varepsilon_m = 10$) versus the separation distance. The curves are computed including : the dipolar moments ($l=1$), all the multipoles up to $l=3$ and $l=5$. The inset is a zoom on the near contact region.

Fig. 8 : (a) Plasmonic thermal conductance at 900 K of a linear chains of 10 copper particles 10 nm radius (ballistic regime) versus the dielectric constant of host material for several separation distances. (b) Plasmonic thermal conductivity at 900 K of a linear chains of 1000 10 nm-radius copper particles (diffusive regime) versus the dielectric constant of host material for several separation distances. (c) Plasmonic thermal conductivity at 900 K of a linear chains of 1000 copper particles in contact versus the dielectric constant of host material for several particle radius. The curves are computed including all the multipoles up to $l=5$.

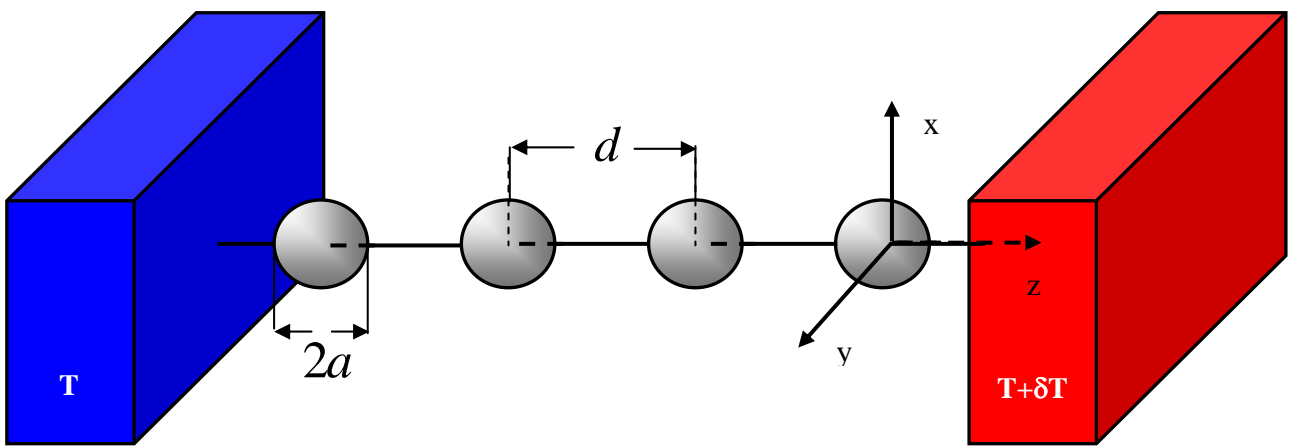

**Figure 1**

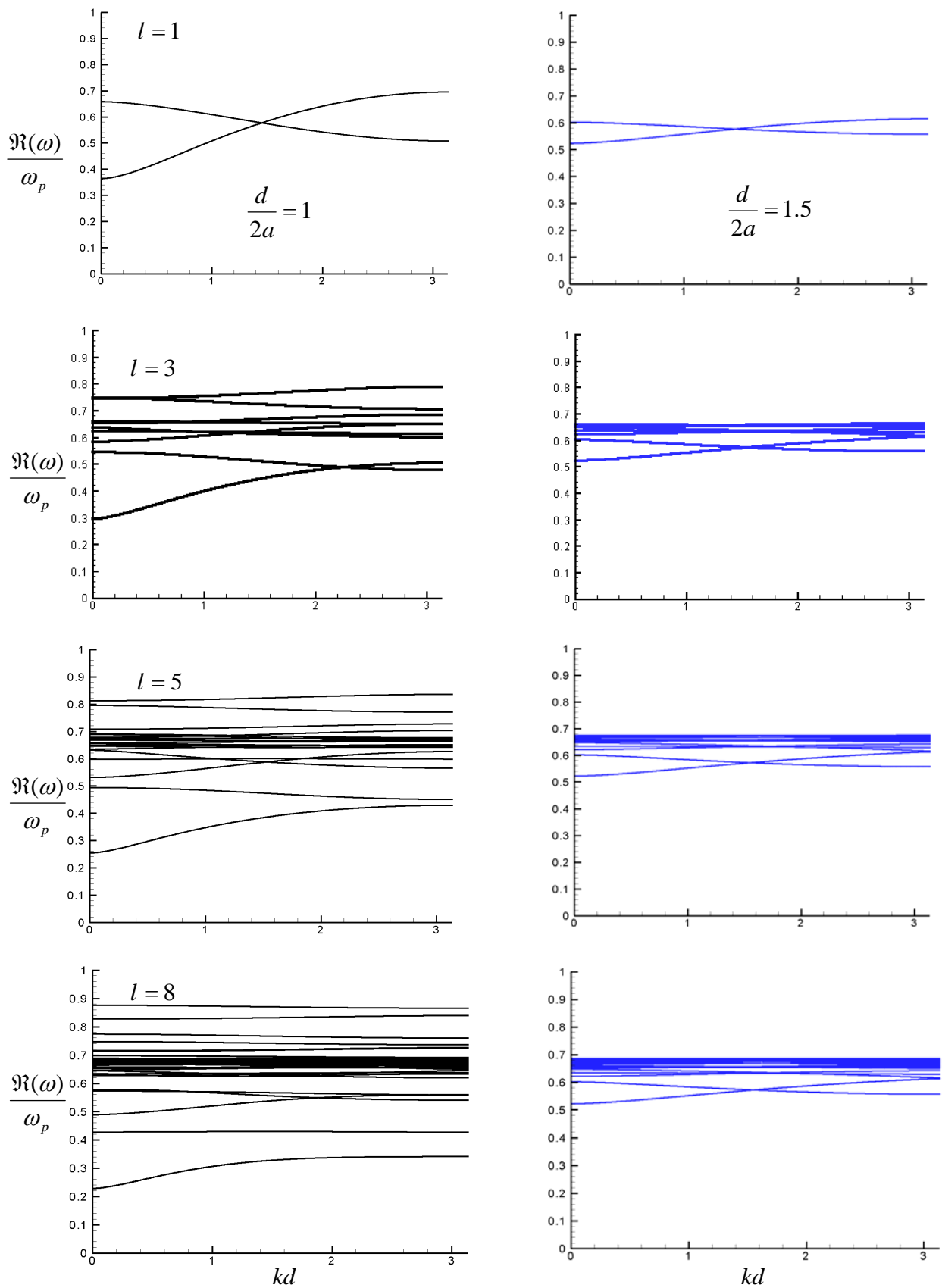

**Figure 2**

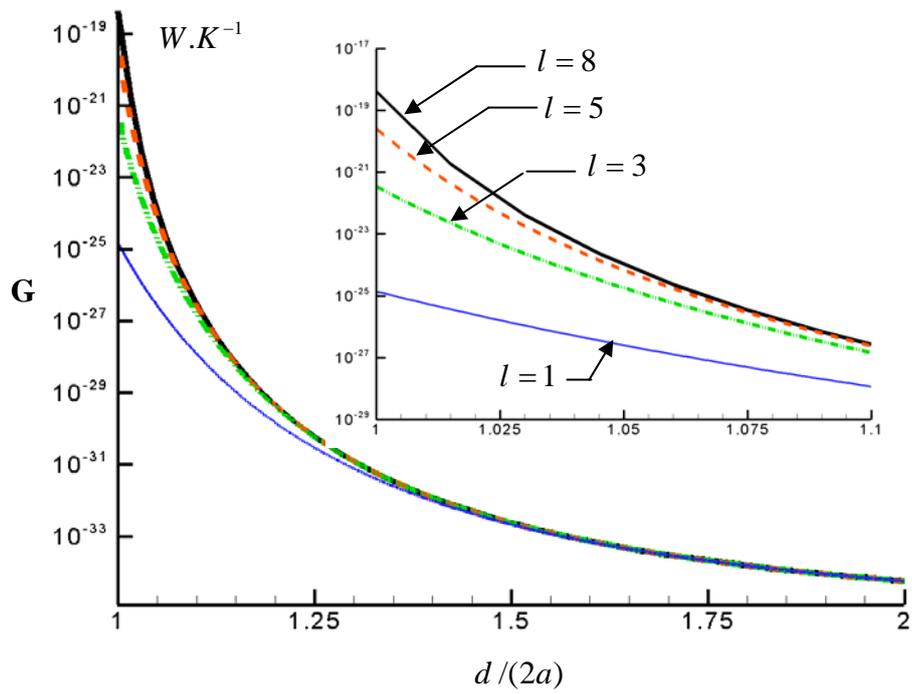

**Figure 3**

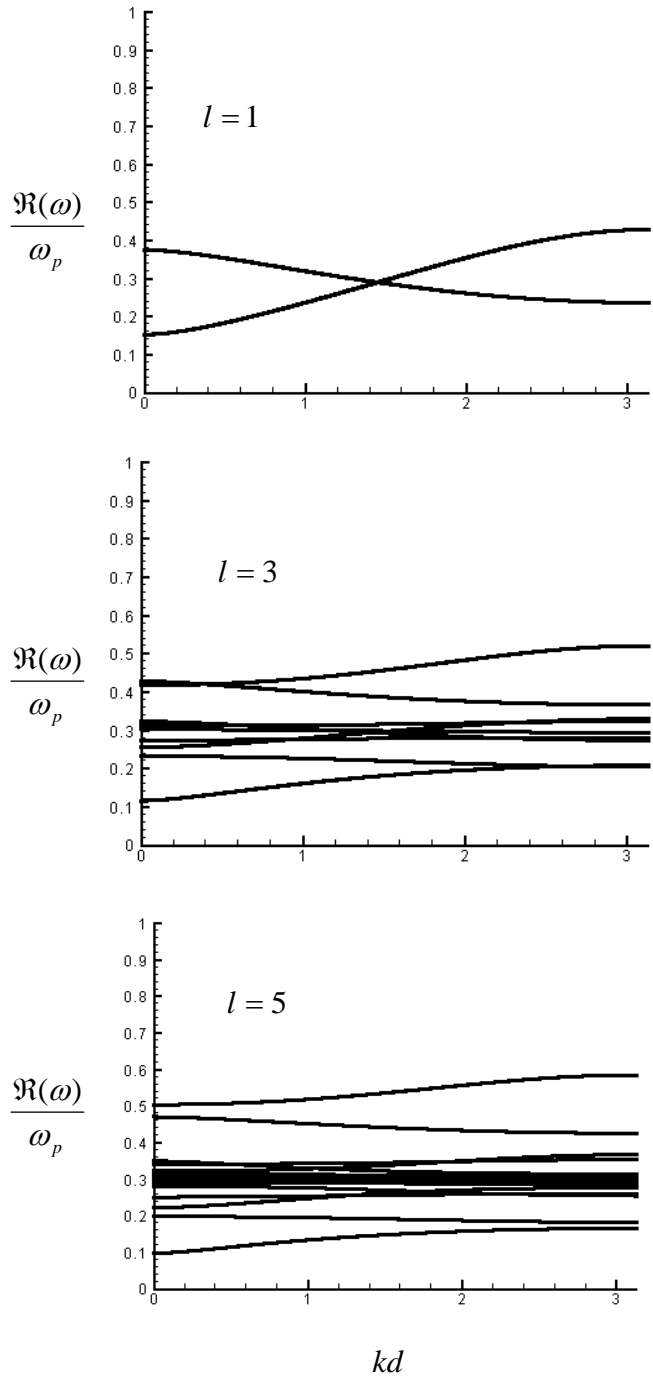

**Figure 4**

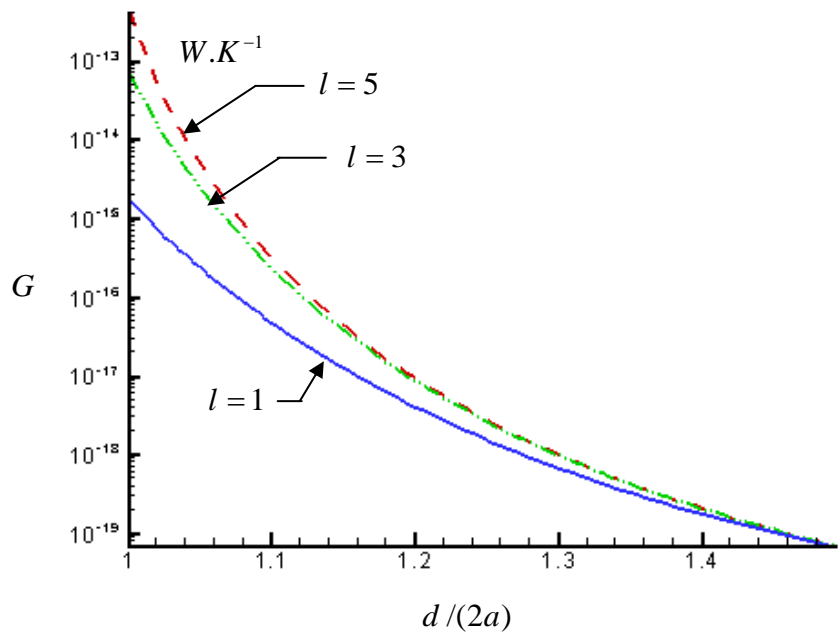

**Figure 5**

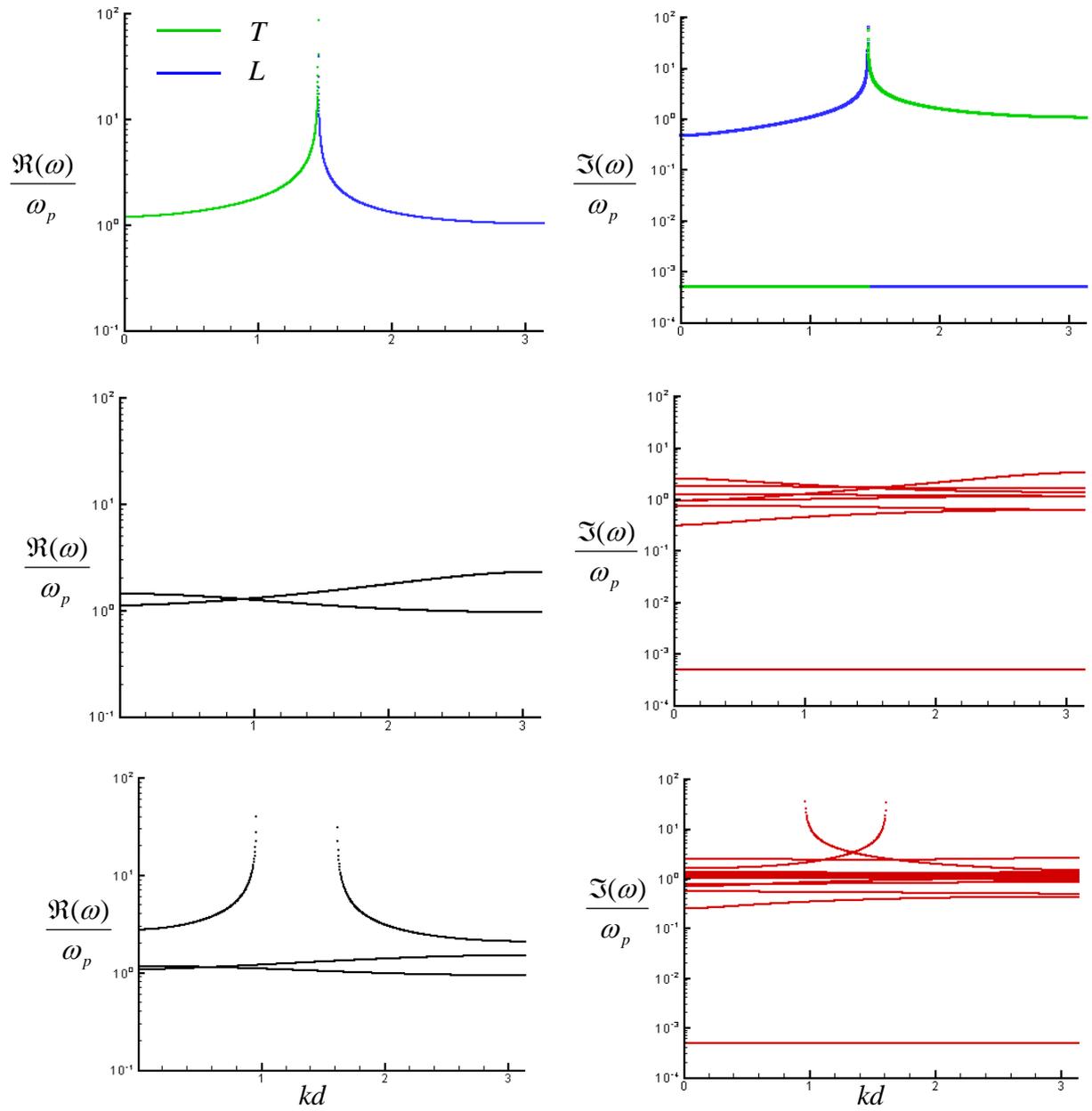

**Figure 6**

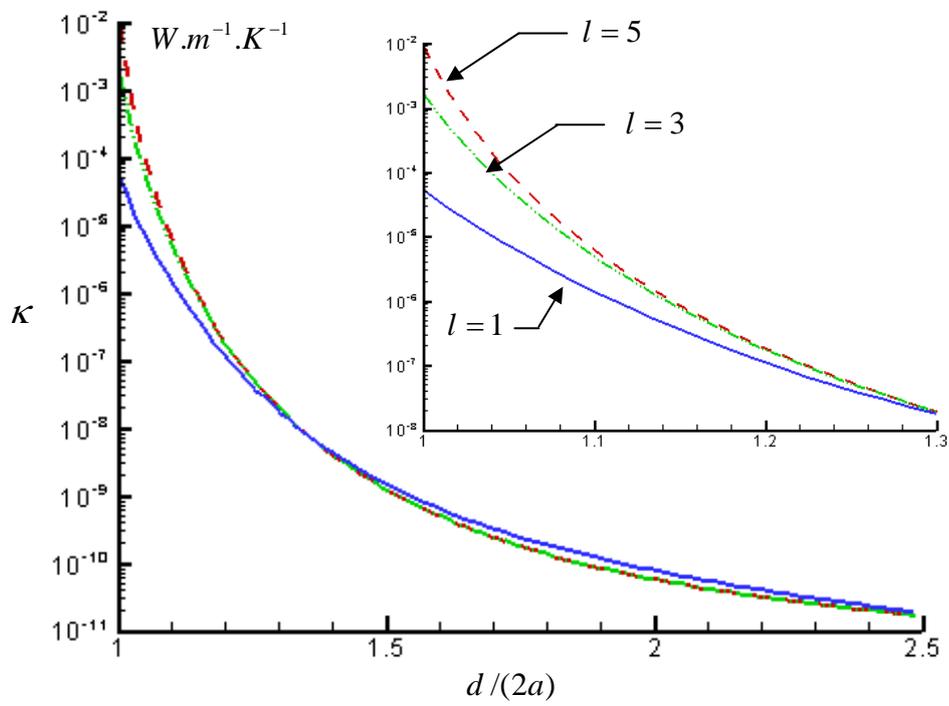

**Figure 7**

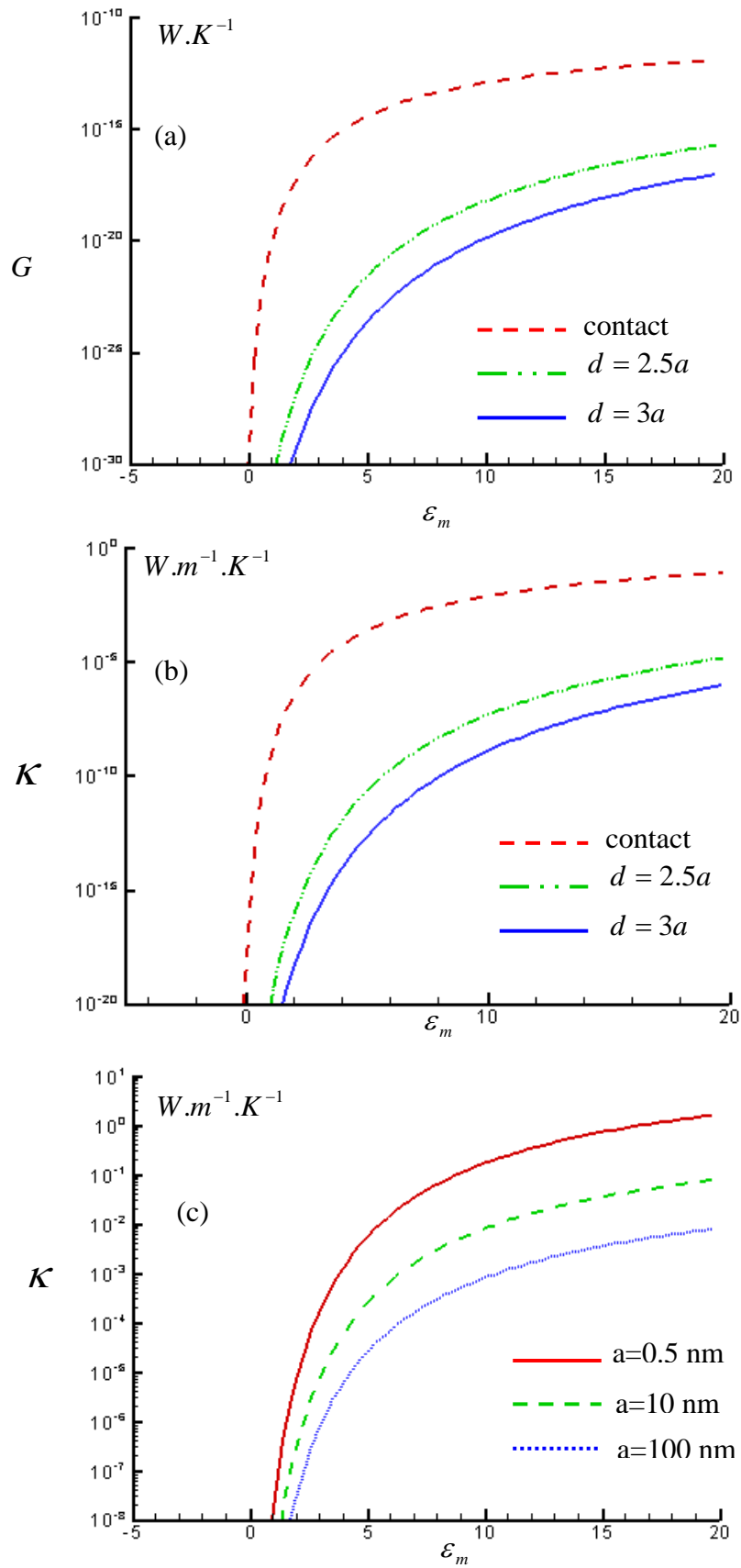

**Figure 8**